\documentclass[epj]{svjour}
\usepackage{graphics}
\begin{document}
\title{Effects on the Non-Relativistic Dynamics of a Charged Particle Interacting with a Chern-Simons Potential}
\author{F. Caruso\inst{1,}\inst{2,}\thanks{\emph{e-mail: francisco.caruso@gmail.com}} \and J.A. Helay\"{e}l-Neto\inst{1,}\thanks{\emph{e-mail: helayel@cbpf.br}} \and J. Martins\inst{2,}\thanks{\emph{e-mail: jordan.martins@uerj.br. The CAPES of Brazil is acknowledged for a fellowship.}} \and Vitor Oguri\inst{2,}\thanks{\emph{e-mail: oguri@uerj.br}}}
%
%
%
\institute{Centro Brasileiro de Pesquisas F\'{\i}sicas, Rua Dr. Xavier Sigaud, 150, 22290-180, Urca, Rio de Janeiro, RJ, Brazil \and Instituto de F\'{\i}sica Armando Dias Tavares, Universidade do Estado do Rio de Janeiro, Rua S\~{a}o Francisco Xavier, 524, 20550-900, Maracan\~{a}, Rio de Janeiro, RJ, Brazil}
%
%


\authorrunning{Caruso, Helay\"el-Neto, Martins, Oguri}
\titlerunning{Non-Relativistic Dynamics with a Chern-Simons Potential}


\abstract{
The hydrogen atom in two dimensions, described by a Schr\"{o}dinger equation with a Chern-Simons potential, is numerically solved. Both its wave functions and eigenvalues were determined for small values of the principal quantum number $n$. The only possible states correspond to $l=0$. How the result depends on the topological mass of the photon is also discussed. In the case $n=1$, the energy of the fundamental state, corresponding to different choice for the photon mass scale, are found to be comprehended in the interval $-3.5 \times 10^{-3}~\mbox{eV} \leq E \leq -9.0 \times 10^{-2}~\mbox{eV}$, corresponding to a mean radius of the electron in the range $ (5.637 \pm 0.005) \times 10^{-8}~\mbox{cm} \leq \langle r \rangle \leq  (48.87 \pm 0.03) \times 10^{-8}~\mbox{cm}$. In any case, the planar atom is found to be very weekly bounded showing some features similar to the Rydberg atoms in three dimensions with a Coulombian interaction.
} 

\maketitle

\section{Introduction}
\label{sec:1}
\label{intro}
Recently, lower-dimensional quantum-mechanical and theoretical field models have acquired a renewed interest and have boosted a tremendous deal of papers over the few recent years, especially in connection with Condensed Matter Systems. Relevant phenomena such as high-$T_c$ superconductivity \cite{Berdnoz,Anderson2}, fractional quantum Hall effect \cite{Klitzing,Girvin} and spin-charge splitting \cite{Tomonaga,Marchetti} in planar systems have motivated an intensive development of models based on principles, fundamental ideas and technicalities of Quantum Field Theory \cite{Tsvelik}. New materials like the genuinely planar graphene monolayers \cite{Novoselov,Geim}, topological insulators \cite{Kane}-\cite{Qi} and topological superconductors \cite{Read}-\cite{Fu} bring to light what we may say is a new state of matter, the so-called topological matter. In many of these phenomena it is well known that the Chern-Simons interaction plays an important theoretical role since its formalism can be used to provide effective field theories capable to describe them and other similar confined systems \cite{Kerler}. Peculiar properties, which are typical of lower-dimensional systems, are revealed in these systems which motivate the real need to apply ordinary and relativistic Quantum Mechanics, methods of Quantum Field Theory and concepts of gauge theories to build up consistent models that correctly describe the observed phenomena.

In the framework of lower-dimensional Physics, more specifically, planar systems, electromagnetic interaction allows that the photon acquires a non-trivial topological mass without any conflict with its inherent gauge symmetry. There emerges then the so-called Maxwell-Chern-Simons Electrodynamics \cite{Deser1,Deser2}, which, as quantum gauge-field theories, display a remarkable ultraviolet behavior and finds applications in the understanding and description of a great diversity of phenomena which are typical of (1+2) space-time dimensions (that is one temporal and two spatial dimensions), such as, for the example, the appearance of particles, the anyons \cite{Leinaas,Wilczek}, which obey an intermediate statistics and are present in many planar systems of interest.


For more details about the physics brought about by the Chern-Simons topological mass term and the emergence of a massive photon in connection with the Meissner Effect, which characterises type-I superconductors \cite{Christiansen}, we point out the detailed works by Dorey and Mavromatos \cite{Dorey1,Dorey2}.

Motivated by the importance and particular aspects of Planar Electrodynamics, we place our question of interest in this frame: namely, we wish to exploit and discuss the influence of space-time dimensions on the electromagnetic interactions and on the properties of particles with electric charge and/or magnetic properties. With this purpose in mind, our paper sets out to investigate the non-relativistic quantum mechanics of an electron in the field of a static point charge considering that the electromagnetic phenomenon is mediated by the (gauge-invariant) massive photon of a Maxwell-Chern-Simons Electrodynamics. By following this path, our true aim is to reassess properties of the non-relativistic hydrogen atom and, so, to compare them with the corresponding properties in other dimensions to really point out the effects of dimensionality on the structure of the hydrogen atom.

The justification to pursue an investigation on the planar hydrogen atom comes originally from the study of the statistical mechanics of the two-dimensional Coulomb gas, as pointed out in the work of ref.~\cite{Belich}. In this article, the authors motivate the study of the planar Schr\"{o}dinger equation to analyze bound-state formation of a binary system of opposite charges. To strengthen the relevance of studying bound states in planar Quantum Mechanics, we quote the papers of Refs.~\cite{Belich-inc1,Costa}, where electron-electron pair states are discussed in connection with superconductivity in different scenarios of planar Electrodynamics. More recently, the work of Ref.~\cite{Fonseca} suggests an interesting possibility to experimentally realize a 2D hydrogen atom in connection with magnetic vortices in topological insulators.

In connection with the physical properties of the electromagnetic interaction in (spatial) two dimensions, we would like to highlight the relevance of the study and calculation of electron-electron bound states in parity-preserving planar systems to describe pair formation in high-$T_c$ cuprate superconductors
\cite{Christiansen,Belich}. The same issue, namely, the possibility of electron-electron pair formation has been investigated for the Maxwell-Chern-Simons-Proca QED(3) \cite{Belich2,Belich-inc1}. Having in mind the peculiarities of the electromagnetic interaction in physically realizable planar systems, we believe that reconsidering the analysis of hydrogen-like atoms in two spatial dimensions is an issue to be further investigated and we focus in the present contribution our attention to this matter.

The outline of our paper is as follows: in Section~\ref{sec:2}, we set up the non-relativistic planar Schr\"{o}dinger equation for an electron in a Maxwell-Chern-Simons potential. Next, in Section~\ref{sec:3}, we carry out the numerical studies of the problem and understand that $l=0$ is the only possible state for the Maxwell-Chern-Simons planar atom. In this $s$-wave state, we work out the values of the binding energy and mean radius of the electron orbit, investigating how those values depends on the choice of the topological mass of the photon involved in the Maxwell-Chern-Simons Electrodynamics. Finally, in Section~\ref{sec:4}, we cast our Concluding Comments.

\section{Hydrogen atom for Chern-Simons potential}
\label{sec:2}

Let us start from the general idea that topologically massive gauge theories are useful to construct models of confined systems. For example, it is known that if one consider a charged particle moving in constant external magnetic and electric fields and if the motion is constrained to be planar and rotationally symmetric, then the vector potential component of the minimal coupling takes the form of a Chern-Simons interaction \cite{Dunne}. Therefore, let us investigate the behavior of a planar hydrogen atom which interaction is dictated by a Chern-Simons potential in (1+2)$D$.

For the classical Maxwell-Chern-Simons Electrodynamics in $(1+2)$ dimensions, the magnetic and the electric fields of a pointlike charge $e$ are given, in the Gaussian unit system, by
$$ B = - \frac{m_{\gamma}c e}{2\pi \hbar}\, K_\circ (m_{\gamma}c r/\hbar); \qquad \vec E = \frac{\hbar}{m_\gamma c} \vec \nabla B = - \vec \nabla \phi$$
where $K_\circ$ is the modified Bessel function and $m_{\gamma}$ is the photon effective topological mass.
The photon topological mass is responsible for the screening of the two-dimensional ($\ln r$)-Coulombian potential, just like the Proca mass term which comes out phenomenologically in the London equation. The difference, however, lies on the gauge invariance: while the Proca mass explicitly breaks down gauge symmetry, the Chern-Simons topological mass reconciles massive photon and gauge invariance. The value of $m_{\gamma}$ is not univocally predicted and, thus, has to be inferred from different physical processes.
It is expected to be in the range $0.1~\mbox{eV} < m_\gamma < 1~\mbox{eV}$ in the case of type II high-temperature superconductors. For conventional superconductors of type I, considering clean elements like Al, In, Sn, Pb and Nb, the measured values for the penetration lengths point to a photon mass parameter in the range $10-20$~eV; for type II conventional superconductors, compounds like Pb-In, Nb-Ti, Nb-N and Pb-Bi indicate that the $m_\gamma$ parameter may take values in the range $2-10$~eV \cite{Orlando}.

From these equations we get the electric potential
$$\phi (r) = \frac{e m_{\gamma}c}{2\pi\hbar}\, K_\circ (m_{\gamma}c r/\hbar)$$

In two dimensions, the Schr\"{o}dinger equation for a massive charged particle (electron), with mass $m$ and charge $e$, interacting with a central external potential $\phi$ is

\begin{equation}
\label{eq:2dim}
\frac{\mbox{d}^{2}u(r)}{\mbox{d}r^{2}} + \frac{2m}{\hbar^{2}}\left\{ E - U(r) - \frac{\hbar^{2}}{2m} \frac{[l^{2} - 1/4]}{r^{2}} \right\} u(r) = 0,
\end{equation}

\noindent where the potential energy is $U(r) = - e \phi(r)$ and the usual radial solution (in $D$ dimensions) is given by
$$R(r) =\displaystyle \frac{u(r)}{r^{(D-1)/2}}.$$

Therefore, the Schr\"{o}dinger equation for a Maxwell-Chern-Simons potential is
\begin{eqnarray}
\label{eq:2dimgeral}
  \nonumber \, \frac{\mbox{d}^{2}u(r)}{\mbox{d}r^{2}} &+& \frac{2m}{\hbar^{2}}\left\{ E +\frac{m_\gamma ce^{2}}{2\pi\hbar}K_{0}\left( \frac{m_\gamma c}{\hbar} r \right) + \right. \\
   \qquad &-& \left. \frac{\hbar^{2}}{2m} \frac{[l^{2} - 1/4]}{r^{2}} \right\} u(r) = 0
\end{eqnarray}


In order to solve numerically Eq.~(\ref{eq:2dimgeral}), it is convenient to put it in a dimensionless form.
Defining Bohr's radius as $a_{B} \equiv \displaystyle\frac{\hbar^{2}}{me^{2}}$, one gets

\begin{eqnarray}
\label{eq:2dimgeral3}
  \nonumber \frac{\mbox{d}^{2}u(r)}{\mbox{d}r^{2}} &+& \frac{1}{a_{B}^{2}}\left\{ \frac{E}{e^{2}/2a_{B}} +\frac{m_\gamma ca_{B}}{\pi\hbar}K_{0}\left( \frac{m_\gamma c}{\hbar} r \right) + \right. \\
   \qquad &-& \left.  a_{B}^{2} \frac{[l^{2} - 1/4]}{r^{2}} \right\} u(r) = 0.
\end{eqnarray}


Introducing the dimensionless variable $x = r/a_{B}$, Eq.~(\ref{eq:2dimgeral3}) can be written as

\begin{eqnarray}
\label{eq:2dimgeral5}
  \nonumber \frac{\mbox{d}^{2}y(x)}{\mbox{d}x^{2}} &+& \left\{ \frac{E}{e^{2}/2a_{B}} +\frac{m_\gamma ca_{B}}{\pi\hbar}K_{0}\left( \frac{m_\gamma ca_{B}}{\hbar} x \right) + \right. \\
   \qquad &-& \left.  \frac{[l^{2} - 1/4]}{x^{2}} \right\} y(x) = 0.
\end{eqnarray}


Finally, expressing $m_\gamma = \lambda m_e$, Eq.~(\ref{eq:2dimgeral5}) can be also written in terms of the fine structure constant, $\alpha$, as

\begin{equation}
\label{eq:estruturafina}
\alpha = \frac{e^{2}}{\hbar c} = \frac{\hbar^{2}}{ma_{B}}\frac{1}{\hbar c} = \frac{\hbar}{mca_{B}} = \frac{1}{137}.
\end{equation}
from which we gets the dimensionless equation
\begin{eqnarray}
\label{eq:2dimfinal}
  \nonumber \frac{\mbox{d}^{2}y(x)}{\mbox{d}x^{2}} &+& \left\{ \frac{E}{e^{2}/2a_{B}} + \frac{\lambda}{\pi\alpha}K_{0}\left( \frac{\lambda x}{\alpha} \right) + \right. \\
   \qquad &-& \left.  \frac{[l^{2} - 1/4]}{x^{2}} \right\} y(x) = 0.
\end{eqnarray}


Thus the problem is reduced to the interaction of a charged particle with an effective potential, $U_{\mbox{\scriptsize{eff}}}(x)$, defined by:

\begin{equation}
\label{eq:potef}
U_{\mbox{\scriptsize{eff}}}(x) = -\frac{\lambda}{\pi\alpha}K_{0}\left( \frac{\lambda x}{\alpha} \right) + \frac{[l^{2} - 1/4]}{x^{2}}.
\end{equation}
which generalizes what was done in Ref.~\cite{Atabek}.

Such an effective potential in two dimensions has a very peculiar behavior, namely, what is usually called ``centrifugal potential'' became indeed attractive when $l=0$. For the Chern-Simons potential, given in Eq.~(\ref{eq:potef}), it is straightforward to see that it is possible to find bound states only when $l=0$.

\begin{figure}
\resizebox{0.50\textwidth}{!}{\includegraphics{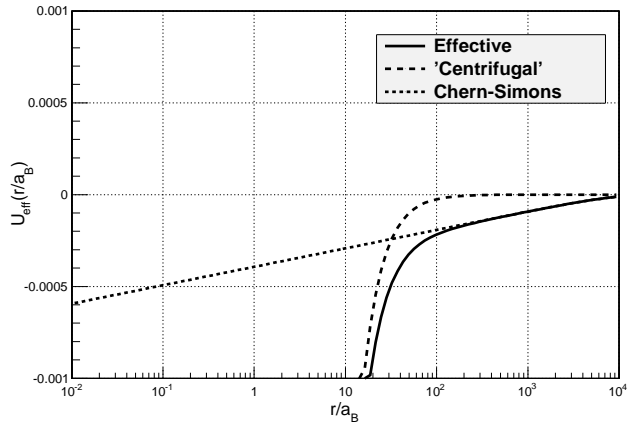}}
\caption{Plot of the effective potential and of each one of its components, for $l = 0$.}
\label{fig:potefL0}
\end{figure}

This effective potential, for $l=0$, and $m_\gamma/m_e = \lambda \simeq 0.2 \times 10^{-5}$, has the behavior as shown in Fig.~\ref{fig:potefL0}.


The eigenvalue equation, Eq.~(\ref{eq:2dimfinal}), can be numerically solved by a slightly modified version of the Numerov Meth\-od \cite{Numerov_1,Numerov_2,Blatt}, as done in Ref.~\cite{Caruso}, in order to include also first derivatives. The algorithm was implemented using $C^{++}$ language. All the calculations and graphics were done by using the CERN/ROOT package.


\section{General results}\label{our_result}
\label{sec:3}

First of all, it is important to say that it is not our intention to carry out a systematic calculation of the hydrogen atom spectra; only few eigenstates and eigenfunctions were determined. Once more, let us stress that for the Maxwell-Chern-Simons potential, the planar structure of the space imposes solutions only with $l=0$. For different values of the angular momentum, the centrifugal potential dominates and the effective potential became positive and continuously decreases.

The second point, already mentioned, is that the result clearly depends on the choice of the topological mass of the photon, $m_{\gamma}$. Thus, for numerical purpose, we have considered three cases: $\lambda= m_{\gamma}/m_e = 0.2 \times 10^{-5}$, $\lambda= 0.2 \times 10^{-4}$ and $\lambda= 0.2 \times 10^{-3}$. For $m_{\gamma} \simeq 0.1~\mbox{eV}$ we could not find a solution within the sensitivity of our numerical calculation. This situation is very close to that where only the `centrifugal term' contributes which gives rise to no bounded state. Conversely, once some energy levels of this kind of planar atom is measured we can straightforwardly set up the scale of $m_{\gamma}$.

\begin{figure}
\resizebox{0.50\textwidth}{!}{\includegraphics{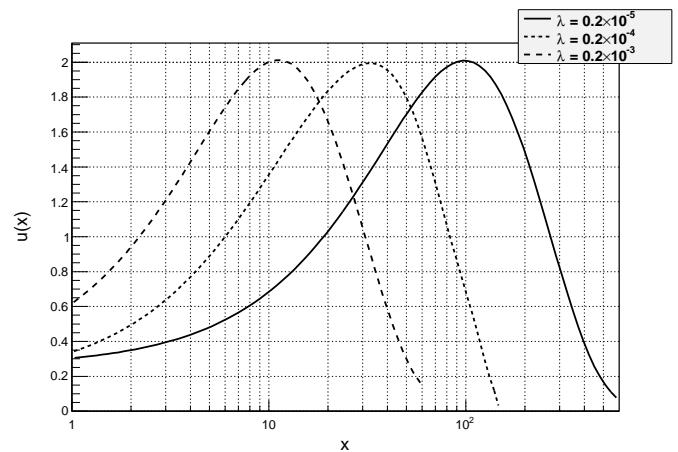}}
\caption{Radial wave function $u_{l=0} (x=r/a_{B})$ for the hydrogen atom in the case $D = 2$ with arbitrary normalization. The three curves correspond to different choices of $m_\gamma$.}
\label{fig:wfL0}
\end{figure}

For the ground state we found numerical solutions for the all probed $\lambda$ scales. The three wavefunctions are shown in Fig.~\ref{fig:wfL0} and the corresponding eigenvalues for each one are shown in Table~\ref{tab:1}.

\begin{table}[h]
 \caption{Dependence of the ground state energy on the parameter $\lambda$.}
 \label{tab:1}
\begin{tabular}{lll}
\hline\noalign{\smallskip}
$\lambda$ & E$_{1}$ $[Ry]$ & E$_{1}$ $[eV]$ \\
\noalign{\smallskip}\hline\noalign{\smallskip}
 $0.2\times10^{-5}$ & -0.00026 & -0.0035\\
 $0.2\times10^{-4}$ & -0.0016 & -0.022 \\
 $0.2\times10^{-3}$ & -0.0067 & -0.09\\
 \noalign{\smallskip}\hline
 \end{tabular}
 \end{table}

\begin{figure}
\resizebox{0.50\textwidth}{!}{\includegraphics{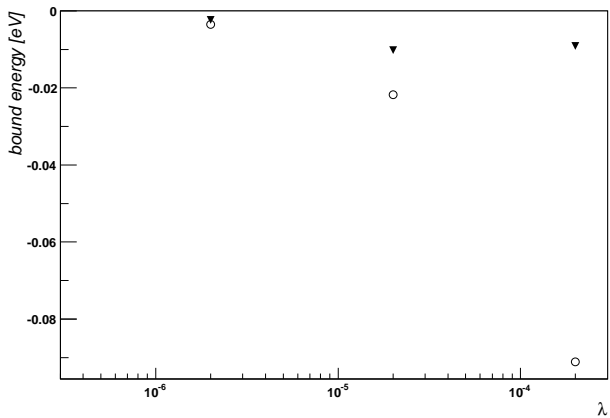}}
\caption{Values of $E_1$ (circle) and $E_2$ (triangle) for different choices of the parameter $\lambda=m_\gamma/m_e$.}
\label{fig:lambda}
\end{figure}

In the case $\lambda= 0.2 \times 10^{-5}$ ($m_\gamma \simeq 1$~eV), we still find a second state ($n=2$, $l=0$) with energy $E_2 = -0.00017$~Ry. For $\lambda= 0.2 \times 10^{-4}$ ($m_\gamma \simeq 10$~eV), two other states are found ($n=2$, $l=0$ and $n$=3, $l=0$) with energies given, respectively, by $E_2 = - 0.00075$~Ry and $E_3= - 0.0004$~Ry.
The dependence of the first two energy levels on the photon topological mass (parameter $\lambda$) is shown in Fig.~\ref{fig:lambda}.

We see that the gap between the first two energy level grows as $\lambda$ grows. An experimental determination of this gap can be used to settle up the $m_\gamma$ scale. 

\begin{figure}
\resizebox{0.50\textwidth}{!}{\includegraphics{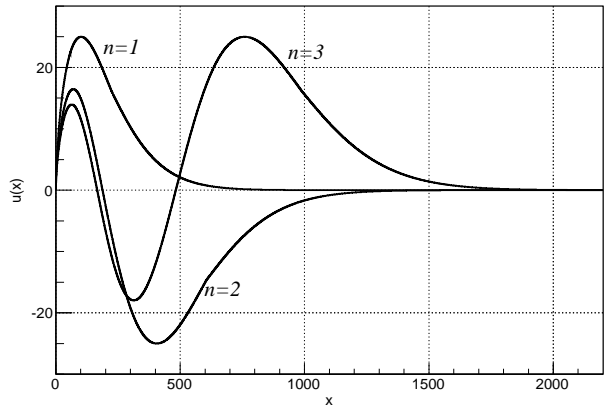}}
\caption{Radial wave functions $u_n(x)$ for the planar hydrogen atom for the states $n=1, 2, 3$ with arbitrary normalization.}
\label{fig:wfn}
\end{figure}

Fig.~\ref{fig:wfn} shows the first three eigenfunctions corresponding to $m_\gamma = 1$~eV.

Once the radial solution is known, the mean value of the electron radius, $\langle r \rangle$, can be evaluated by solving the expression (with $D=2$)

\begin{equation}
\label{eq:rmedio}
 \langle r \rangle = \frac{\int r^{D} \cdot R_{l=0}^{2} (r) \mbox{d}r}{\int r^{D-1} \cdot R_{l=0}^{2} (r) \mbox{d}r} =
   a_{B}\frac{\int x \cdot y_{l=0}^{2} (x) \mbox{d}x}{\int y_{l=0}^{2} (x) \mbox{d}x}
\end{equation}
which was done numerically adopting the Monte Carlo Rejection Method \cite{Cowan,James}.

\begin{figure}
\resizebox{0.50\textwidth}{!}{\includegraphics{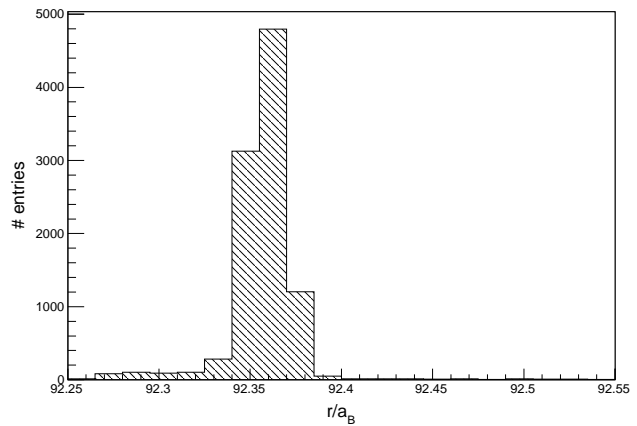}}
\caption{Histogram generated by the Rejection Monte Carlo Method in order to estimate the mean value of the electronic radius in the ground state of the planar hydrogen atom, corresponding to the choice $m_{\gamma} = 1~\mbox{eV}$.}
\label{fig:medioR}
\end{figure}

For the ground state, and considering $m_{\gamma} = 1~\mbox{eV}$, the result for the mean radius is $\langle x \rangle = \langle r \rangle/{a_{B}} = (92.35 \pm 0.06)$, which corresponds to the statistic treatment of the histogram given in  Fig.~\ref{fig:medioR}. Using the value of Bohr's radius, $a_{B} =  0.52917720859(36) \times 10^{-8} ~\mbox{cm}$ \cite{PDG} (in Gaussian unit system), the mean radius of the electron in the ground state of a planar hydrogen atom described by the Schr\"{o}dinger equation with a Chern-Simons potential, is $\langle r \rangle = (48.87 \pm 0.03) \times 10^{-8} ~\mbox{cm}$. This result is two orders of magnitude greater than the atomic radius of the hydrogen atom in three dimensions ($\sim 10^{-8} ~\mbox{cm}$). The values of the orbital radius for different choices of $m_{\gamma}$ ($\lambda$) are shown in Table~\ref{tab:2}.

\begin{table}
 \caption{Mean values of $r/a_{B}$ by varying $\lambda$ (or $m_\gamma$).}
 \label{tab:2}
\begin{tabular}{ll}
\hline\noalign{\smallskip}
$\lambda$ & $\langle r \rangle / a_{B}$ \\
\noalign{\smallskip}\hline\noalign{\smallskip}
 $0.2\times10^{-5}$ & $92.35 \pm 0.06$ \\
 $0.2\times10^{-4}$ & $25.59 \pm 0.01$ \\
$0.2\times10^{-3}$ & $10.653 \pm 0.009$ \\
 \noalign{\smallskip}\hline
 \end{tabular}
 \end{table}

 These values can be compared to the results of the same mean value for hydrogen atomic radius in higher dimensional spaces \cite{Caruso}, from the number of dimensions varying from 3 to 10 (remember that for $D=4$ there is no solution for the hydrogen atom) leading to the conclusion that in two dimensions the orbital radius of the electron in the ground state is by far the biggest one.


We would like to conclude this Section making some comments concerning the possibility of experimental verification of our results. In general, it is true that many of the predictions based on Chern-Simons theory applied to planar condensed-matter systems are hard to be experimentally proved. However, so far the planar hydrogen atom is concerned one can be more optimistic. In fact, in Ref.~\cite{Baxter}, a suitable experimental arrangement able to give us an experimental verification of the Chern-Simons feature of fractional angular momentum of an atomic dipole in constant electric and magnetic fields is proposed. Such a possibility is discussed in the context of cold Rydberg atoms \cite{Gallagher}. That cold Rydberg atoms play an interesting role of realizing analogous of Chern-Simons theory is stressed also in Refs.~\cite{Zhang1,Zhang2}. All these ideas depend on the general feature that an atomic dipole of a cold Rydberg atom could be arranged in appropriate external electric and magnetic fields, so that the motion of the dipole is constrained to be planar and rotationally symmetric. We shall turn back to this point in the next Section. in any way, according to Ref.~\cite{Zhang2}, its is possible to prepare the planar atomic dipole in its energy ground state, which will be a convenient configuration for testing our results.

%
%

\section{Final Comments}
\label{sec:4}

Let us compare our result with the well known features of the hydrogen atom in three dimensions. Remember that the ground state of this 3$D$ simple atom with such dimensionality is bound by 1~Ry, and the orbital radius of the electron is one Bohr's radius, 1~$a_B$. In contrast, the $n=10$ state, which can be considered as a relatively low Rydberg state,  has a binding energy of $0.01$~Ry and a radius of 100~$a_B$. Now, if we compare our result for the 2$D$ ground state, bounded by the energy $E_1~=~0.00026$~Ry, with the three-dimensional analogous energy $E_n \sim 1/n^2$, we see that our prediction for the ground state energy of a planar atom corresponds to the energy of a 3$D$ atom with the principal quantum number $n=62$. For this state, the calculated mean atomic radius is
$\langle r \rangle  \simeq 92 a_B$, instead of the value 100~$a_B$ found in three-dimensions. Notice that the binding energy of the fundamental state of hydrogen atom in two dimensions (3.5~meV) is comparable to one tenth of thermal energies ($\simeq~40$~meV). Therefore, one should expect that planar atoms with one electron might be a useful system with which to probe the interaction of an atom and the vacuum \cite{Gallagher}. Also the possibility of having giant molecules \cite{Bend},  \cite{Farooqi}, \cite{Boisseau}, could be investigated.


Let us now turn back to the fact that one way to experimentally get planar hydrogen atoms is by starting from cold Rydberg atoms. Notice that the calculation made in this paper does not depend anyway on the experimental process such planar atomic dipoles are obtained. Therefore, the result we get (namely, that due to the (1+2)-Chern-Simons interaction the planar hydrogen atom should have characteristics similar to those of Rydberg atoms in three spatial dimensions) indeed teaches us that even in its ground state the atomic electron is still very far from the proton ($\langle r \rangle  \simeq 92 a_B$). Thus, this original result suggests that the dimensional reduction imposed to a Rydberg hydrogen atom by a strong cooling process still maintain it feature of being weekly bounded, with a large value for $\langle r \rangle$, as a Rydberg atom is expected to be. However, as already stressed, such a configuration does not depend on how this 2$D$ atom is builded.

In two space dimensions, new aspects of the electromagnetic interactions could be considered that may enrich our study of the hydrogen atom and introduce new corrections to its binding energy and mean radius. Among these, we should consider a non-minimal coupling \cite{Paul},\cite{Nobre},\cite{Paschoal}  through the dual of the electromagnetic field strength, along with the minimal coupling with the potential in the gauge covariant derivative. This is property specific of (1+2)$D$. Also, it would be of interest, for the sake of computing effects due to the magnetic properties of the electron, to re-analyze the magnetic moment and the gyromagnetic ratio (the $g$-factor) of the electron in a planar scenario to get its influence on the corrections to the quantities we have numerically computed in Section~\ref{sec:3}.

   It is our immediate purpose to go through this direction and we shall be soon reporting on that elsewhere.




\end{document}